\pdfoutput=1

\documentclass[11pt]{article}

 \usepackage{acl}

\usepackage{times}
\usepackage{latexsym}
\usepackage{multirow}
\usepackage{amsmath}
\usepackage{bm}
\usepackage{graphicx}
\usepackage{tabularray}
\usepackage{tabularx}
\usepackage{amssymb}
\usepackage{enumitem}
\usepackage{color}
\usepackage{ragged2e}
\usepackage{booktabs}
\usepackage{hhline}
\usepackage{rotating}
\usepackage[ruled, vlined]{algorithm2e}
\usepackage{xcolor}
\usepackage{makecell} 

\SetCommentSty{mycommfont}

\usepackage[normalem]{ulem}
\usepackage{array}
\usepackage{arydshln}

\usepackage[T1]{fontenc}

\usepackage[utf8]{inputenc}

\usepackage{microtype}

\graphicspath{ {./Figures/} }
\title{IterCQR: Iterative Conversational Query Reformulation with Retrieval Guidance}

\author{
Yunah Jang$^{1}$ Kang-il Lee$^{1}$ Hyunkyung Bae$^{2}$ Hwanhee Lee$^{3\dagger}$ Kyomin Jung$^{1,4\dagger}$ \\
    $^{1}$Dept. of ECE, Seoul National University $  $
    $^{2}$LG AI Research $  $
    $^{3}$Chung-Ang University $  $\\
    $^{4}$IPAI, Seoul National University $  $\\
    \texttt{\{vn2209, 4bkang, kjung\}@snu.ac.kr}\\ \texttt{hkbae@lgresearch.ai}, \texttt{hwanheelee@cau.ac.kr}\\
}
\begin{document}
\maketitle
\footnotetext{\textsuperscript{$\dagger$}Corresponding authors.}

\begin{abstract}
Conversational search aims to retrieve passages containing essential information to answer queries in a multi-turn conversation. 
In conversational search, reformulating context-dependent conversational queries into stand-alone forms is imperative to effectively utilize off-the-shelf retrievers. 
Previous methodologies for conversational query reformulation frequently depend on human-annotated rewrites.
However, these manually crafted queries often result in sub-optimal retrieval performance and require high collection costs.
To address these challenges, we propose \textbf{Iter}ative \textbf{C}onversational \textbf{Q}uery \textbf{R}eformulation \textbf{(IterCQR)}, a methodology that conducts query reformulation without relying on human rewrites. 
IterCQR iteratively trains the conversational query reformulation (CQR) model by directly leveraging information retrieval (IR) signals as a reward.
Our IterCQR training guides the CQR model such that generated queries contain necessary information from the previous dialogue context.
Our proposed method shows state-of-the-art performance on two widely-used datasets, demonstrating its effectiveness on both sparse and dense retrievers. 
Moreover, IterCQR exhibits superior performance in challenging settings such as generalization on unseen datasets and low-resource scenarios.\footnote{Code and datasets are available at \url{https://github.com/YunahJang/IterCQR}}


\end{abstract}

\section{Introduction}

\begin{figure}[th]
\centering
\includegraphics[width=0.95\columnwidth]{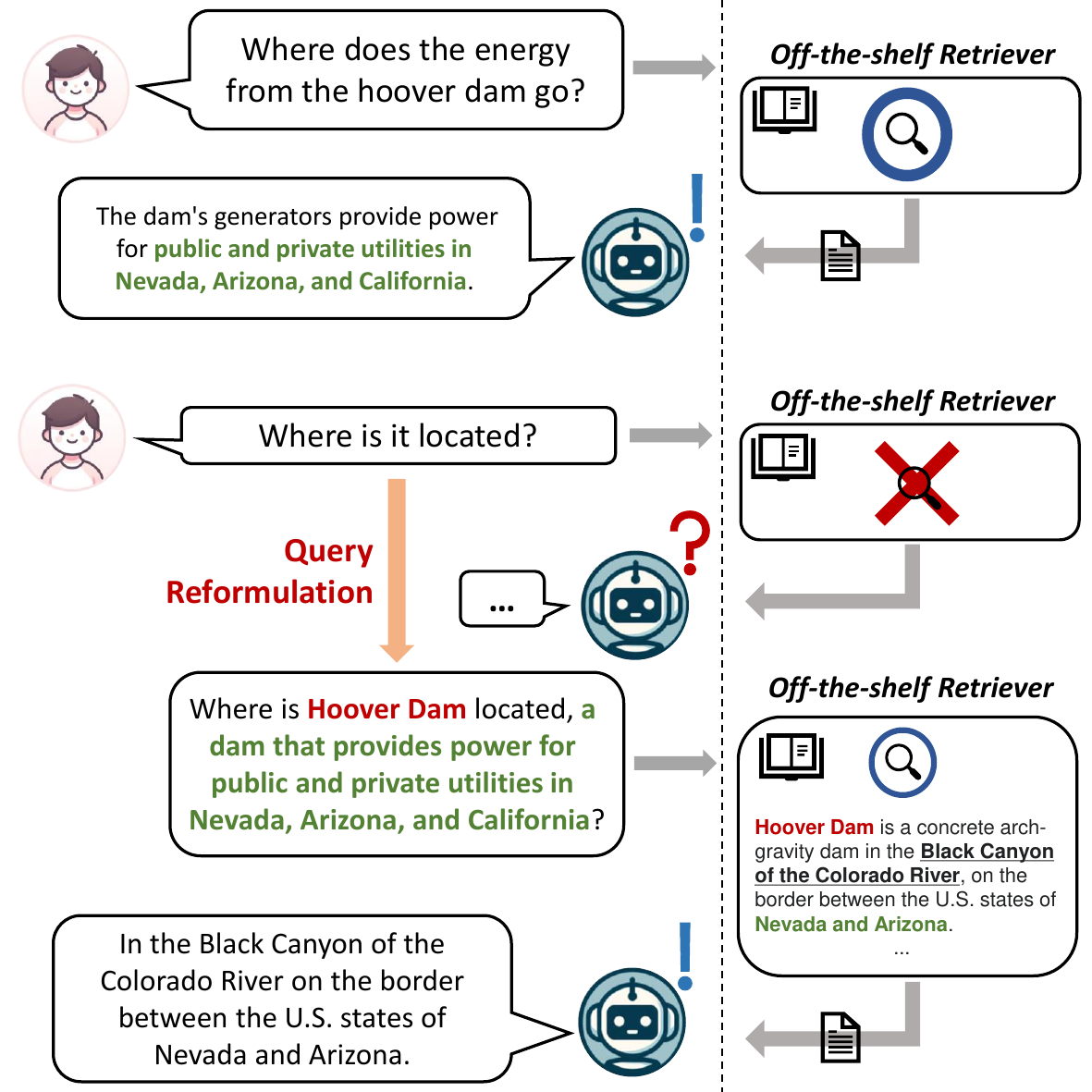}
\caption{In the CQA task, the user's queries are dependent on the previous dialogue context. CQR task reformulates conversational queries into stand-alone queries, which are then fed into the off-the-shelf retrievers.}
\vspace{-5mm}
\label{fig:task}
\end{figure}
In the conversational question answering (CQA) task, questions and answers are exchanged in a multi-turn conversation.  
As a component of CQA task, conversational search aims to retrieve passages that contain the necessary information to answer the current query within the conversation \cite{qrecc, topiocqa}.

Owing to the conversational setting, queries in CQA suffer from a high dependency on the previous conversation context, as shown in Figure \ref{fig:task}, introducing challenges such as omissions, ambiguity, and coreference \cite{llm_cqr,pseudopassage}.
Therefore, in conversational search, conversational queries cannot be directly used as inputs for off-the-shelf retrievers trained on non-contextual queries.

One possible strategy for mitigating this challenge is to train retrievers to comprehend long dialogue context \cite{yu_denseretriever, kim_denseretriever, lin_denseretriever}. 
However, this method results in substantial cost in retraining retrievers to handle long inputs. 
As an alternative method, researchers have explored conversational query reformulation (CQR) that reformulates conversational queries into stand-alone questions, which enables the utility of off-the-shelf retrievers \cite{convgqr, quretec}.

CQR methods can be categorized into rewriting and expansion.
Most prior research trains models for query rewriting using human-annotated gold labels \cite{quretec, common-ground}. 
However, these manually crafted queries often yield sub-optimal performance \cite{contextualized, conqrr} in addition to requiring costly and time-consuming collection process. 
Human labels tend to focus on de-contextualizing queries based on human subjective judgment, which does not always align with passage retrieval performance \cite{conqrr}.

To address human rewrite's sub-optimality, ongoing research explores various expansion methods, such as potential answer expansion \cite{convgqr}, and classifying previously mentioned entities for expansion \cite{common-ground, quretec}.
However, these expansions are not directly optimized for retrieval signals.
Also, the existence of separate expansion and rewriting models requires additional training steps, storage, and two rounds of inference for a single query.

In this paper, we propose an \textbf{Iter}ative \textbf{C}onversational \textbf{Q}uery \textbf{R}eformulation (IterCQR) model, that iteratively performs query reformulation without using human rewrites.
Since ground truth labels are unavailable, our approach employs an iterative framework to alternate between generating candidate queries and optimizing CQR model with their IR signals as a reward.

\begin{figure*}[th!]
\centering
\includegraphics[width=\textwidth]{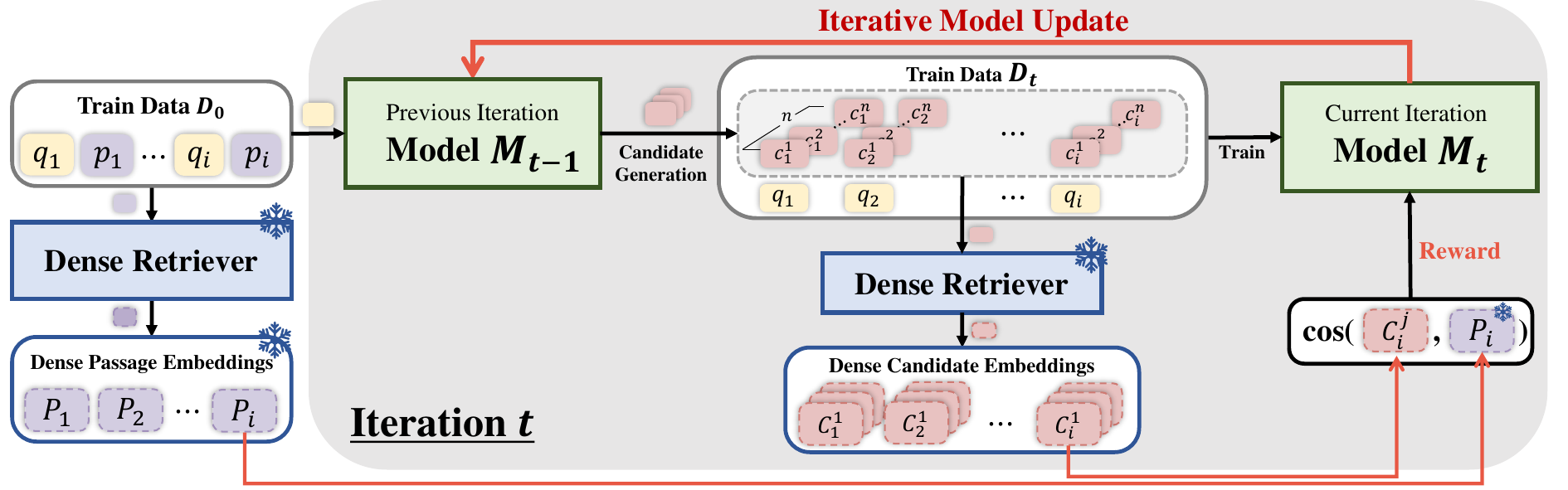}
\caption{Overview of IterCQR. IterCQR trains on the candidates generated by the previous iteration model. We define reward as the cosine similarity between the frozen dense passage embeddings and dense candidate embeddings.}
\vspace{-3mm}
\label{fig:overview}
\end{figure*}

For the initialization of IterCQR, we leverage LLMs' ability to create an initial rewritten query dataset for training CQR model.
After training IterCQR with the initial dataset, we iteratively train the model on generated query candidates through the Minimum Bayes Risk (MBR) \cite{mbr} training method and Top-1 candidate selection.
In the training process, we integrate IR signal by defining the reward value as the cosine similarity between reformulated queries and ground-truth passages.
After the iterative training process for IterCQR, we employ the final iteration model to reformulate queries, which are then utilized as inputs for off-the-shelf retrievers.

IterCQR achieves state-of-the-art performance on two widely used CQA datasets, the TopiOCQA \cite{topiocqa} and QReCC \cite{qrecc} datasets.
We also show that IterCQR exhibits superior performance in various scenarios, such as generalization on unseen datasets and low-resource settings. 
Through a quantitative analysis of iterative reformulation, we experimentally demonstrate that IterCQR generates summary expansion from the preceding context as the iteration progresses. 
This expansion contributes to an improvement in retrieval performance, demonstrating the ability of IterCQR to generate retriever-friendly queries.\\ 
\vspace{-3mm}
The main contributions of this work are as follows: 
\vspace{-3mm}
\begin{itemize}[noitemsep,leftmargin=*]
\item We propose IterCQR which iteratively trains a conversational query reformulation model without human label while utilizing off-the-shelf retrievers.
\item IterCQR exhibits state-of-the-art performance on both the TopiOCQA and QReCC datsets.
\item IterCQR shows outstanding performance in challenging scenarios including generalization on unseen datasets and low-resource settings.
\end{itemize}




\section{Related Works}

\subsection{Conversational Query Reformulation}
CQR focuses on improving conversational search performance by rewriting and expanding user queries in a conversational context.
In contrast to other conversational search methods, the reformulated queries in CQR can be directly utilized as input to off-the-shelf retrievers without fine-tuning.

Previous studies have addressed CQR by using human-rewritten queries or query expansion methods \cite{lin2020conversational, voskarides, shi-fewshot}.
However, human rewrites have been reported to be sub-optimal \cite{contextualized, conqrr}, and expansion methods require a separate model for term selection \cite{voskarides,kumar-callan-2020-making} or potential answer generation \cite{convgqr}. 

To address these shortcomings, \textsc{Con}QRR \cite{conqrr} employs Self-Critical Sequence Training (SCST) \cite{scst} to directly optimize the query rewriting model to the retriever. 
More recently, ConvGQR \cite{convgqr} integrates query rewriting and expansion to further enhance retrieval performance; however, it requires two separate models for rewriting and expansion, which hampers efficiency in both the training and inference processes.
Furthermore, these approaches require expensive human-rewritten queries.
Recent works show that LLMs are capable of reformulating queries \cite{llm_qr}, including conversational queries \cite{llm_cqr, pseudopassage, promptagator}.
But LLM-generated queries also require further optimization for retrievers similar to human-rewritten queries.

Our work proposes an alternative approach of directly optimizing the CQR model to the retriever, using only gold passage annotation.
Both IterCQR and \textsc{Con}QRR utilize a reward defined by the IR signal. However, \textsc{Con}QRR uses a binary reward function based on the retrieval results of BM25, whereas we employ cosine similarity with the gold passage, which is a real-valued reward.
Additionally, \textsc{Con}QRR generates candidates at each time step, which means the target sequences change dynamically throughout the training. 
On the other hand, IterCQR repetitively learns over a static target sequence, providing a consistent and efficient training signal.
We employ Minimum Bayes Risk (MBR) training \cite{mbr, iterative} and maximum likelihood training based on the top-1 candidate to effectively learn without human annotated queries.

\subsection{Iterative Learning in NLG tasks}
The conventional method for training natural language generation (NLG) models uses human oracles, which are costly and time-consuming to collect; moreover, quality control during the collection is challenging. 
Therefore, research has been focused on learning without human supervision through iteratively enhancing the quality of the training dataset.
Many works on weakly supervised QA and semantic parsing revolve around such iterative refinement paradigms, such as iterative search \cite{iterative}, ambiguous learning \cite{ambiguous}, and hard EM approach \cite{min-etal-2019-discrete}.
In task-oriented dialogue, \citet{gptcritic} propose to iteratively update the training set with self-generated samples.
In this paper, we apply an iterative framework on the CQR task, which iteratively optimizes updated training samples with retrieval guidance.

\section{Method}
\subsection{Problem Definition}
Conversational search aims to retrieve relevant passages containing rich information that can answer the current conversational query \(q\) within a CQA system.
To achieve this goal, conversational queries are reformulated so that we can utilize an off-the-shelf retriever \(R\), which has been trained on non-conversational question-answering data.

We train a query reformulation model \(M\) to rewrite and expand the original query $q$ based on the previous history context \(H\) to generate a de-contextualized query \(q^{*}\).
Training input for the query reformulation model \(M\) on turn $k$ is \(\{q_k,H_{k-1}\}\), concatenation of current query and the history context, where history context \(H_{k-1}\) is a consecutive sequence of previous queries and answers in reversed order.
The reformulated query \(q^{*}_k\) from \(M\) is then used as input to the off-the-shelf retriever \(R\), which retrieves a ranked list of the top-$k$ relevant passages.

\subsection{IterCQR}
IterCQR utilizes the iterative setting to train CQR model without relying on human-rewritten queries.
Specifically, our CQR model utilizes IR signals to generate the optimal query for retrieval tasks.

\SetKwInput{KwModel}{Model} 
\SetKwInput{KwInput}{Input} 
\begin{algorithm}[hbt!]
\small 
\DontPrintSemicolon
\SetAlgoLined
\SetNoFillComment
\caption{IterCQR}\label{alg:IterCQR}
\KwInput{ Conversational query for \(k^{th}\) turn \(q_k\), Previous history context \(H_{k-1}\)}
\KwData{Train Data \(D\) without human-rewritten query}
\KwModel{CQR Model $M_t$ for iteration $t$ }
\KwResult{Reformulated query for \(k^{th}\) turn \(q^{*}_k\), Query candidate for \(k^{th}\) turn \(c_{k}^j\)}
\For {iteration $t =0..T$ } { 
    \For{$q^k, H_{k-1} \in D $}{
    \uIf (\tcp*[f]{Initialize $D_0$}) { $t=0$ } { 
    $ q^{*}_k= LLM(q_k,H_{k-1}) $\; 
    $ D_0 \gets D_0 \cup  q^{*}_k $\;}
    \Else (\tcp*[f]{Generate $D_t$}){  
    \For {$j=0..n$}{
    $ c_{k}^{j}= M_{t-1}(q_k,H_{k-1}) $\; 
    $ D_t \gets D_t \cup c_{k}^j $\;}}}
       \uIf(\tcp*[f]{Train}){ $t=0$ }
       { $\mathcal{L} \gets {\mathcal{L}_{NLL}}$ with target $q^{*}_k $\; } 
       \uElseIf{$t\leq \tau$ \small{(exploration factor)}}
       { $ \mathcal{L} \gets {\mathcal{L}_{MBR}}$ with $n$ candidates $c_{k}^{j}$\;} 
       \Else{ Select Top-1 Candidate \(c_i^{top}\)  \;
        $ \mathcal{L} \gets \mathcal{L}_{NLL}$ with target $c_i^{top}$\; }
        Train \(M_t\) with \(D_t\) to minimize \( \mathcal{L} \)\;
}
\end{algorithm}

We describe the training process of IterCQR in Algorithm \ref{alg:IterCQR}.
We first initialize IterCQR by training on the initial dataset \(D_0\), which contains queries rewritten by LLM. 
After initializing the model, we go through the iterative process shown in Figure \ref{fig:overview}.
During iteration \(t\), we first utilize the previous iteration \(t-1\) model \(M_{t-1}\) to generate \(n\) candidate queries for each instance in the training set \(D\).
Subsequently, this newly created training dataset, \(D_{t}\), becomes the training data for the \(M_{t}\) model.
In this iterative process, starting from \(M_{1}\), candidates generated by the previous iteration model become the targets for training the current iteration model. 

IterCQR leverages the cosine similarity between the dense embedding of the candidate query and gold passage to guide the CQR model to generate retriever-friendly queries.
This reward prioritizes candidates with the most relevant semantic representation to the gold passage.
Furthermore, for enhanced learning efficiency, we define exploration factor \(\tau\) to balance the exploration and exploitation in the training phase.
At the iterations less than or equal to \(\tau\), we employ the MBR training algorithm to facilitate exploration, followed by an exploitation phase using Top-1 candidate selection approach.

\subsubsection{Data Initialization with LLM}
We construct the initial dataset \(D_0\) by utilizing LLM, \texttt{gpt-3.5-turbo}, to rewrite the queries.\footnote{The prompts used for initializing \(D_0\) is shown the Appendix \ref{prompt}.}
The initial model \(M_0\) is trained on $D_0$ with negative log-likelihood loss.
Although LLMs show great abilities in various tasks, they still exhibit limitations in conversational query reformulation, considering that LLMs are not optimized for retrievers. 
Hence, starting from Iteration 1, we optimize IterCQR using IR signals.

\subsubsection{MBR Training} 
In the early stages of the training, iterations less than or equal to \(\tau\), \(D_{t}\) contains diverse candidates that are suitable for exploration through the application of the Minimum Bayes Risk (MBR) training method.
By employing all $n$ candidates, the model can learn not only from the high-probability candidates but also from candidates with lower probability values.
The MBR training algorithm seeks to minimize the expected value of a cost function $\mathcal{C}$ between input $x$ and candidate $y$. 
\newcommand{\E}{\mathbb E} 
\begin{equation} \label{eq:mbr_orig}
   \min_{\theta}\sum_{i=1}^{N} \E_{\tilde{p}(y_i|x_i;\theta)}\mathcal{C}(x_i, y_i)
\end{equation}

Here, we approximate the expectation using the re-normalized probabilities of the candidates obtained through beam search, denoted as \(\tilde{p}\).
To apply the MBR training algorithm for CQR model, we define a reward function $\mathcal{R}$ instead of the cost function $\mathcal{C}$.
The MBR training loss is formulated to minimize the negative MBR term, as expressed in Equation \ref{eq:mbr_loss}:
\begin{equation}\label{eq:mbr_loss}
\begin{split}
 &{\mathcal{L}_{MBR}}=-\sum_{i=1}^{N}\mathbb{E}_{\tilde{p}(c_i|q_i,H_{k-1};\theta)}\mathcal{R}(C_i, P_{i}) \\
 & =-\sum_{i=1}^{N}\sum_{j=1}^{n}{\tilde{\mathbb{P}}(c^{j}_{i}|q_i,H_{k-1};\theta)} \cdot\mathcal{R}(C_{i}^{j}, P_{i}).
\end{split}
\end{equation}


To employ the IR signal in the reward value, we set the reward function as the cosine similarity between the dense embedding of candidate query \(C_i\) and the dense embedding of the ground-truth passage \(P_i\), as shown in Equation \ref{eq:reward}. 
We generate both passage and candidate query embeddings using the frozen encoder of the dense retriever.

\begin{equation}\label{eq:reward}
    \mathcal{R}(C_i, P_i) = \cos(C_i, P_i)
\end{equation}
We observe that most reward values fall within a limited range, which could hamper providing fine-grained learning signals. 
To address this issue, we apply min-max normalization to scale the reward distribution into a range of 0 to 1.


By utilizing reward signals for all \(n\) candidate queries, the MBR approach enables learning across a diverse set of queries, ultimately resulting in significant improvements in the retrieval performance.
After conducting query exploration through MBR iteration, the model proceeds to perform query exploitation by Top-1 candidate selection to effectively utilize the acquired knowledge.

\subsubsection{Top-1 Candidate Selection} 
Following the standardization and enhancement of candidate quality through MBR training, we perform exploitation through Top-1 candidate selection.
The exploitation objective aligns with exploration, reformulating conversational queries with retriever guidance.
In this step, we select the top-1 candidate \(c^{top}\) among \(n\) candidates as the target for the training.

In this process, the criteria for Top-1 candidate selection also rely on the cosine similarity between candidate embedding and the dense embedding of the gold passage.
This selection criterion ensures alignment with the IR signal. 
We use the negative log-likelihood loss in the exploitation step to maximize the likelihood of generating top-1 candidate \(c_i^{top}\) as Equation \ref{eq:nll_loss}:

\begin{equation}\label{eq:nll_loss}
    \mathcal{L}_{NLL}=-\sum_{i=1}^{N} \log (\mathbb{P}(c_i^{top}|q_i,H_{k-1})).
\end{equation}

Through this two-step approach, we aim to prevent queries from diverging too far from the semantic space of existing queries compared to training only with the MBR training algorithm for exploration.
It also reduces computational complexity by eliminating the step of recalculating probabilities for all \(n\) candidates.


\subsection{Retriever Models}
We test IterCQR on both dense and sparse retrievers.
Following the previous works \cite{llm_cqr,mao2022curriculum, convgqr, yu_denseretriever}, we use BM25 for sparse retriever and ANCE \cite{ance} for the dense retriever.
We generate dense embedding of gold passages and candidates using the ANCE dense retriever finetuned on MS MARCO \cite{msmarco}. 
We store these embeddings and re-use them for the entire training and inference steps because the dense retriever is frozen from the beginning. 
We also use the ANCE model for encoding candidate queries' dense embeddings to calculate reward.

\section{Experiments}
\paragraph{Dataset} We train and evaluate our model on QReCC \cite{qrecc} and TopiOCQA \cite{topiocqa}. 
Both datasets consist of conversational queries and corresponding gold answers paired for each turn. 
Notably, TopiOCQA includes topic labels determined based on Wikipedia documents.
\noindent

\begin{table*}[th]
\centering\resizebox{\textwidth}{!}{
\begin{tabular}{@{\extracolsep{8pt}}cl|cccc cccc@{}}
\toprule [1.5pt]

\multirow{2}{*}{\textbf{Type}} & \multirow{2}{*}{\textbf{Method}} & \multicolumn{4}{c}{\textbf{TopiOCQA}} & \multicolumn{4}{c}{\textbf{QReCC}} \\ 
\cline{3-6}
\cline{7-10}
 &  & \textbf{MRR} & \textbf{NDCG@3} & \textbf{R@10} & \textbf{R@100} & \textbf{MRR} & \textbf{NDCG@3} & \textbf{R@10} & \textbf{R@100} \\ 
\hline
\multirow{8}{*}{\begin{tabular}[c]{@{}c@{}}\textbf{Dense}\\\textbf{(ANCE)}\end{tabular}}
 & Raw & 0.041 & 0.038 & 7.5 & 13.8& 0.102 & 0.093 & 15.7 & 22.7\\
 & Initial & 0.178 & 0.168 & 32.6 & 47.7 & 0.358 & 0.330 & 55.7 & 74.1 \\
 & GPT2QR & 0.126 & 0.120 & 22.0 & 33.1 & 0.339 & 0.309 & 53.1 & 72.9 \\
 & QuReTeC & 0.112 & 0.105 & 20.2 & 34.4 & 0.350 & 0.326 & 55.0 & 70.9 \\
 & T5QR & 0.230 & 0.222 & 37.6 & 54.4 & 0.345 & 0.318 & 53.1 & 72.8 \\
 & \textsc{Con}QRR & - & - & - & - & 0.418 & - & 65.1 & 84.7 \\
 & ConvGQR & 0.256 & 0.243 & 41.8 & 58.8 & 0.420 & 0.391 & 63.5 & 81.8 \\
 & \textsc{Edi}RCS & - & - & - & - & 0.421 & - & \textbf{65.6} & \textbf{85.3} \\
 & \textbf{IterCQR} & \textbf{0.263} & \textbf{0.251} & \textbf{42.6} & \textbf{62.0} & \textbf{0.429} & \textbf{0.402} & 65.5 & 84.1 \\ 
\cline{2-10}
 & Human-Rewrite & - & - & - & - & 0.384 & 0.356 & 58.6 & 78.1 \\ 
\cline{1-10}
\cline{1-10}
\multirow{8}{*}{\begin{tabular}[c]{@{}c@{}}\textbf{Sparse}\\\textbf{(BM25)}\end{tabular}} 
& Raw & 0.021 & 0.018 & 4.0 & 9.2 & 0.065 & 0.055 & 11.1 & 21.5 \\
 & Initial & 0.132 & 0.115 & 25.2 & 47.3 & 0.322 & 0.290 & 51.8 & 81.2 \\
 & GPT2QR & 0.062 & 0.053 & 12.4 & 26.4 & 0.304 & 0.279 & 50.5 & 82.3 \\
 & QuReTeC & 0.085 & 0.073 & 16.0 & 31.3 & 0.340 & 0.305 & 55.5 & 86.0 \\
 & T5QR & 0.113 & 0.098 & 22.1 & 44.7 & 0.334 & 0.302 & 53.8 & 86.1 \\
 & \textsc{Con}QRR & - & - & - & - & 0.383 & - & 60.1 & 88.9 \\
 & ConvGQR & 0.124 & 0.107 & 23.8 & 45.6 & 0.441 & 0.410 & \textbf{64.4} & 88.0 \\
& \textsc{Edi}RCS & - & - & - & - & 0.412 & - & 62.7 & \textbf{90.2} \\
 & \textbf{IterCQR} & \textbf{0.165} & \textbf{0.149} & \textbf{29.3} & \textbf{54.1} & \textbf{0.467} & \textbf{0.441} & \textbf{64.4} & 85.5 \\ 
\cline{2-10}
 & Human-Rewrite & - & - & - & - & 0.397 & 0.362 & 62.5 & 98.5 \\
\bottomrule [1.5pt]
\end{tabular}}
  \caption{Performance of IterCQR on TopiOCQA and QReCC dataset using dense and sparse retriever. We utilize ANCE for the dense retriever and BM25 for the sparse retriever. \textbf{Bold} letters indicate the best performance of reported results; human-rewrite is excluded in this comparison. We only report the human-rewrite performance of QRECC since TopiOCQA doesn't have human annotation. Note that \textsc{Con}QRR used DualEncoder for dense retriever instead of ANCE.}
  \label{table:main_results}
\end{table*}

\begin{table}[t]
\centering
\resizebox{\columnwidth}{!}{%
\renewcommand{\arraystretch}{1.2}
\begin{tabular}{cccccc}
\toprule [1.5pt]
\multicolumn{6}{c}{\textbf{TopiOCQA Test}} \\ \hline
\multirow{2}{*}{\textbf{Test}} &
  \multicolumn{1}{c|}{\multirow{2}{*}{\textbf{Method}}} &
  \multicolumn{2}{c|}{\textbf{Dense}} &
  \multicolumn{2}{c}{\textbf{Sparse}} \\
 &
  \multicolumn{1}{c|}{} &
  \multicolumn{1}{c}{\textbf{MRR}} &
  \multicolumn{1}{c|}{\textbf{NDCG@3}} &
  \multicolumn{1}{c}{\textbf{MRR}} &
  \multicolumn{1}{c}{\textbf{NDCG@3}} \\ \hline
\textbf{ID} &
  \multicolumn{1}{c|}{\textbf{ConvGQR}} 
& 0.256 &  \multicolumn{1}{c|}{0.243} & 0.124 & 0.107
   \\ 
   \hline
\textbf{OOD} &
  \multicolumn{1}{c|}{\textbf{IterCQR}}
  & 0.178 &  \multicolumn{1}{c|}{0.164}& 0.137 & 0.122
   \\ 
\hline\hline
\cline{1-6}
\cline{1-6}
\cline{1-6}

\multicolumn{6}{c}{\textbf{QReCC Test}} \\ \hline
\multirow{2}{*}{\textbf{Test}} &
  \multicolumn{1}{c|}{\multirow{2}{*}{\textbf{Method}}} &
  \multicolumn{2}{c|}{\textbf{Dense}} &
  \multicolumn{2}{c}{\textbf{Sparse}} \\
 &
  \multicolumn{1}{c|}{} &
  \multicolumn{1}{c}{\textbf{MRR}} &
  \multicolumn{1}{c|}{\textbf{NDCG@3}} &
  \multicolumn{1}{c}{\textbf{MRR}} &
  \multicolumn{1}{c}{\textbf{NDCG@3}} \\ \hline
\multirow{2}{*}{\textbf{ID}} &
  \multicolumn{1}{c|}{\textbf{\textsc{Con}QRR}} 
  & 0.418 &  \multicolumn{1}{c|}{-} & 0.383 & -  \\
 &
  \multicolumn{1}{c|}{\textbf{ConvGQR}} 
  & 0.420 &  \multicolumn{1}{c|}{0.391} & 0.441 & 0.410 \\ 
 \hline
\textbf{OOD} &
  \multicolumn{1}{c|}{\textbf{IterCQR}} 
  & 0.401 &  \multicolumn{1}{c|}{0.374} & 0.449 & 0.424
   \\ 
   \bottomrule [1.5pt]
\end{tabular}%
}
    \caption{ IterCQR performance on unseen datasets. \textsc{Con}QRR and ConvGQR are evaluated in an in-domain (ID) setting, while IterCQR is evaluated in an out-of-domain (OOD) setting. }
    \label{table:zero_shot3}
\vspace{-5mm}
\end{table}

\paragraph{Evaluation Metrics} Our model evaluates retrieval performance on commonly used metrics, such as mean reciprocal rank (MRR), NDCG@3, Recall@10, and Recall@100, following previous works \cite{conqrr,convgqr, qrecc}. 
We utilized the pytrec\_eval tool \cite{pytrec}, as ConvGQR, to calculate these metrics in the subsequent experiments. 
\noindent
\paragraph{Baselines} We compare our IterCQR model with seven baseline models: (1) \textbf{Raw}: This baseline represents the results obtained when using the original query of the data as input.\footnote{Raw results for TopiOCQA are the result of the model trained on QReCC since TopiOCQA does not have a human rewrite.}
(2) \textbf{Initial}: The model \(M_0\) trained by the initial dataset \(D_0\) generated by \texttt{gpt-3.5-turbo}.
(3) \textbf{GPT2QR (Transformer++)} \cite{qrecc}: GPT-2 (medium) \cite{gpt2} based QR model introduced in QReCC as a powerful baseline.
(4)  \textbf{QuReTeC} \cite{quretec}: In this approach, query resolution is treated as a binary classification problem and trained to determine whether to include terms from the previous turn in the current query. 
(5) \textbf{T5QR} \cite{lin2020conversational}: Query reformulation model built upon the T5-base model \cite{t5}.
(6) \textbf{\textsc{Con}QRR} \cite{conqrr}: Leveraging reinforcement learning, \textsc{Con}QRR use retriever signals to generate queries optimized for the retriever.
(7) \textbf{ConvGQR} \cite{convgqr}: ConvGQR achieves strong performance through the integration of rewrite and potential answer expansion. The potential answer expansion model is trained using the gold answers from the dataset.
(8) \textbf{\textsc{Edi}RCS}\cite{edircs}: \textsc{Edi}RCS is a text editing based CQR model that selects tokens to rewrite from the previous dialogue and generates few new tokens. 


We directly report the results from ConvGQR paper for the baselines (RAW, GPT2QR, T5QR, and ConvGQR), \textsc{Con}QRR results from the \textsc{Con}QRR paper, and \textsc{Edi}RCS from the \textsc{Edi}RCS paper. 
\noindent

\paragraph{Implementation Details} 
We use T5-base as a backbone of the CQR model, and use ANCE dense retriever \cite{ance} for the passage encoder which is kept frozen throughout all training iterations.
For all iterations, the number of candidates is set to 10.
We set \(\tau\) as 1 for both the TopiOCQA and QReCC datasets.
For the selection of the \(\tau\) value, refer to \ref{ablation}.
We use Adam optimizer with a learning rate of 1e-5, a batch size of 8, and a maximum query length of 32.
See more details in Appendix \ref{implementation}.

\subsection{Main Results}
We present the results of the IterCQR model trained on TopiOCQA and QReCC, respectively, in Table \ref{table:main_results}.
Although other baseline models were trained or utilized human rewrites for training, IterCQR demonstrates superior performance even without using a human rewrite.

Notably, IterCQR outperforms the second-best performing model, ConvGQR, with a significant improvement in the results of the dense retriever on TopiOCQA.
Although the reward function of IterCQR is defined in terms of dense passage embeddings, the model exhibits a significant performance in the sparse retriever. 
In fact, our model surpasses the retrieval performance of directly using human rewrites.

\subsection{Generalization on Unseen Datasets}
In this section, we show the IterCQR's generalization ability on unseen datasets. We train IterCQR on TopiOCQA and evaluate on QReCC test set and vice versa. The results of these experiments are presented in Table \ref{table:zero_shot3}.

IterCQR, which was solely trained on QReCC and tested in an out-of-domain(OOD) setting for the TopiOCQA test set, outperforms the in-domain(ID) model in sparse retrieval results, across all evaluation metrics.
QReCC-trained IterCQR outperforms TopiOCQA-trained ConvGQR on the TopiOCQA test set in sparse retrieval performance.
Evaluating on the QReCC test set, the TopiOCQA-trained IterCQR shows comparable performance to the ID setting models for the dense passage retriever.
Notably, in the case of sparse retrieval performance on the QReCC test set, despite being tested in an OOD setting, TopiOCQA-trained IterCQR outperforms QReCC-trained ConvGQR and \textsc{Con}QRR, both of which follow the ID setting.
These experiments underscore IterCQR's strong generalization capabilities across diverse datasets.




\subsection{Low-resource Setting}
To evaluate the performance of IterCQR in a low-resource scenario, we demonstrate the results of models trained on 20\% and 50\% of the entire TopiOCQA train set in Table \ref{table:low_resource}.
Remarkably, even when utilizing only 50\% of the full TopiOCQA training data, the models achieve comparable performance to those trained with the entire dataset.
Furthermore, compared to ConvGQR, the second best-performing model in the main results in Table \ref{table:main_results}, both the 20\% and 50\% models surpass ConvGQR on sparse retrieval results and perform comparably in dense retrieval results. 
This observation shows that IterCQR is effective in low-resource scenarios, consistently exhibiting state-of-the-art performance even when trained with a limited amount of training data.
\begin{table}[t]
\resizebox{\columnwidth}{!}{%
\begin{tabular}{@{\extracolsep{5pt}}cc|cccc@{}}
\toprule[1.5pt]
\multirow{2}{*}{\textbf{Type}} & \multirow{2}{*}{\textbf{Data}} & \multicolumn{4}{c}{TopiOCQA}  \\ 
\cline{3-6}
 &  & MRR & NDCG@3 & R@10 & R@100  \\ 
\hline
\multirow{3}{*}{Dense} & 20\% & 0.204 & 0.189 & 35.4 & 55.6 \\
 & 50\% & 0.252 & 0.242 & 41.6 & 58.8  \\
 & 100\% & \textbf{0.263} & \textbf{0.251} & \textbf{42.6} & \textbf{62.0} \\ 
\hline
\multirow{3}{*}{Sparse} & 20\% & 0.144 & 0.128 & 25.5 & 51.7 \\
 & 50\% & 0.162 & 0.145 & 28.7 & 52.9  \\
 & 100\% & \textbf{0.165} & \textbf{0.149} & \textbf{29.3} & \textbf{54.1}  \\
\bottomrule[1.5pt]
\end{tabular}%
}
  \caption{ Performance of IterCQR in a  low resource scenario. We train IterCQR using 20\%, 50\%, and 100\% of the TopiOCQA train dataset.}
  \label{table:low_resource}
\end{table}

\subsection{Ablation Study}  \label{ablation}
\begin{figure}[t]
\centering
\includegraphics[width=\columnwidth]{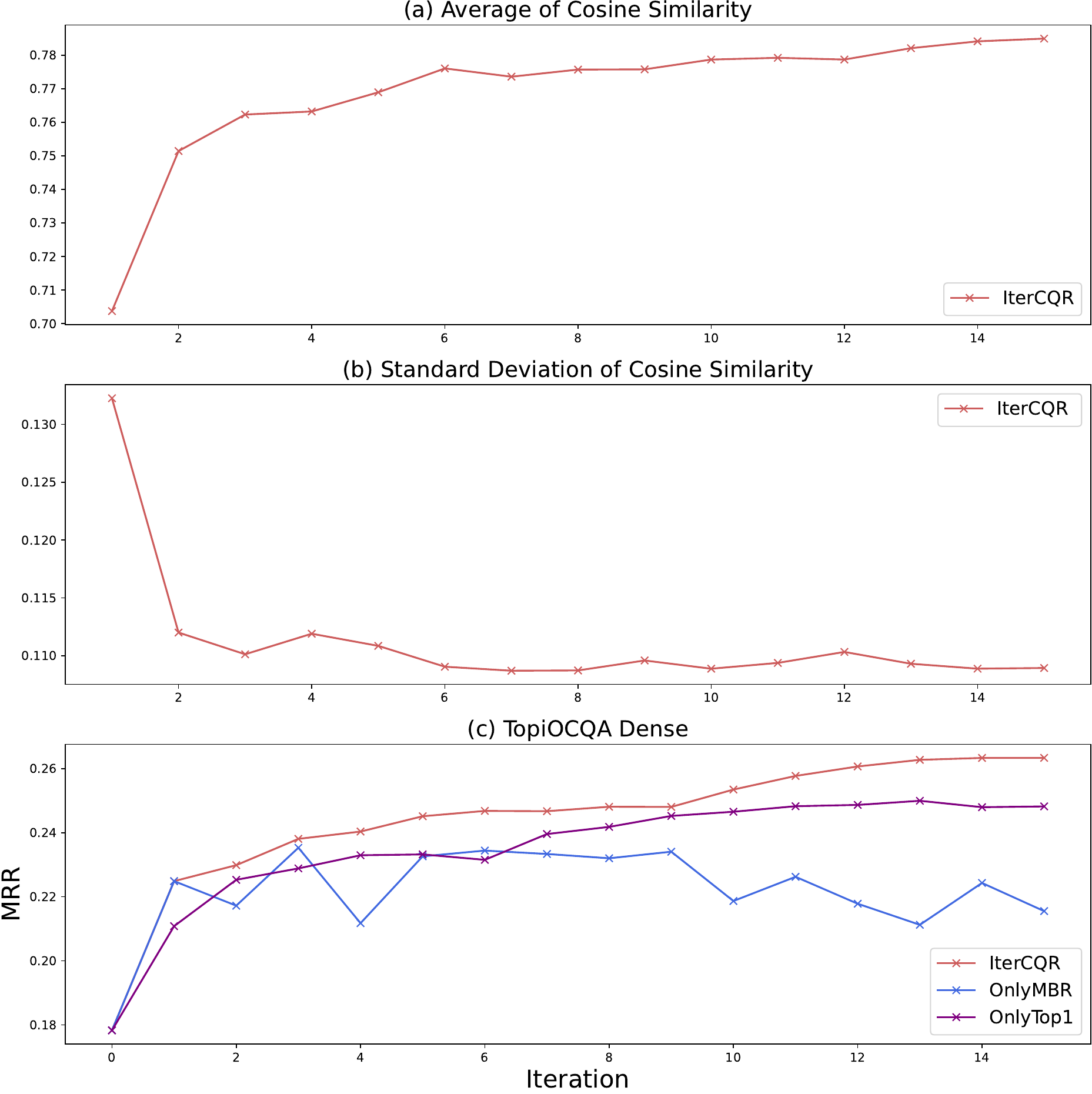}
\caption{OnlyMBR is the model trained only with the MBR algorithm, and OnlyTop1 is trained only with the Top-1 candidate selection.  }
\label{fig:onlymbr}
\vspace{-5mm}
\end{figure}
\begin{table}[t]
\centering
\resizebox{\columnwidth}{!}{%
\begin{tabular}{@{\extracolsep{5pt}}cc|cccc@{}}
\toprule[1.5pt]
\multirow{2}{*}{\textbf{Type}} & \multirow{2}{*}{\textbf{Method}} & \multicolumn{4}{c}{TopiOCQA} \\ 
\cline{3-6}
 &  & MRR & NDCG@3 & R@10 & R@100 \\ 
\hline
\multirow{3}{*}{Dense} & IterCQR & \textbf{\textbf{0.263}} & \textbf{\textbf{0.251}} & \textbf{\textbf{42.6}} & \textbf{\textbf{62.0}} \\
 & OnlyMBR  & 0.216 & 0.204 & 35.6 & 53.6 \\
 & OnlyTop1 & 0.248 & 0.234 & 41.6 & 61.1 \\ 
\hline
\multirow{3}{*}{Sparse} & IterCQR & \textbf{0.165} & \textbf{0.149} & \textbf{29.3} & \textbf{54.1} \\
 & OnlyMBR & 0.111 & 0.099 & 20.0 & 52.9 \\
 & OnlyTop1 & 0.150 & 0.134 & 26.1 & 48.1 \\
\bottomrule[1.5pt]
\end{tabular}%
}
  \caption{ Retrieval performance of IterCQR, OnlyMBR, and OnlyTop1 on TopiOCQA dataset. }
  \vspace{-3mm}
  \label{table:ablation}
\end{table}

In this work, we have presented a two-step training approach; MBR algorithm for exploration, and Top-1 candidate selection for exploitation. 
We conduct an ablation study on the core components of training IterCQR. 

In Table \ref{table:ablation}, it is illustrated that both exploration and exploitation are required; models trained with either one of the components display noticeable drops across metrics tested with both dense and sparse retrievers. This clearly indicates that IterCQR requires two-step training approach, where the model explores-and-exploits the query space. 




We argue the superiority of the two-step training approach derives from the fact that the MBR training method is particularly effective when a diverse range of candidate qualities exists.
After adequate training with MBR, most of the $n$ candidates exhibit high reward values, leading to an unstable training process because of the tendency to penalize candidates with relatively lower reward values, even if the candidates possess good quality.
We observe that, in Figure \ref{fig:onlymbr} (a), the average cosine similarities of candidates exhibit a noticeable increase after the iteration with MBR training method, whereas the standard deviation decreases significantly as in Figure \ref{fig:onlymbr} (b).

Furthermore, in Figure \ref{fig:onlymbr} (c) we observe that IterCQR consistently enhances dense retrieval performance, while OnlyMBR, the model trained with only the MBR training algorithm, exhibits unstable learning.
Moreover, the OnlyTop1 model, trained solely with Top-1 candidate selection, results in slower learning and a lower performance saturation point than IterCQR.
Therefore, it is advisable to initially utilize MBR for exploration, and once the variance in cosine similarity values among candidates has decreased, switch to the exploitation with Top-1 candidate selection to achieve more stable learning and facilitate an efficient query search.
Based on these experimental results, we set the \(\tau\) value to 1, processing MBR training for a single iteration.
See Appendix \ref{ablation_query} for generated queries of IterCQR, OnlyMBR, and OnlyTop1.

\section{Analysis}
We analyze the effect of iterative rewriting on IterCQR by each iteration.
For each iteration, we measure the retrieval performance, token length of the rewritten queries, token overlap with the historical context, and token overlap with the gold passage.

\subsection{Effect on Retrieval Performance}
\begin{figure}[t]
\centering
\includegraphics[width=0.9\columnwidth]{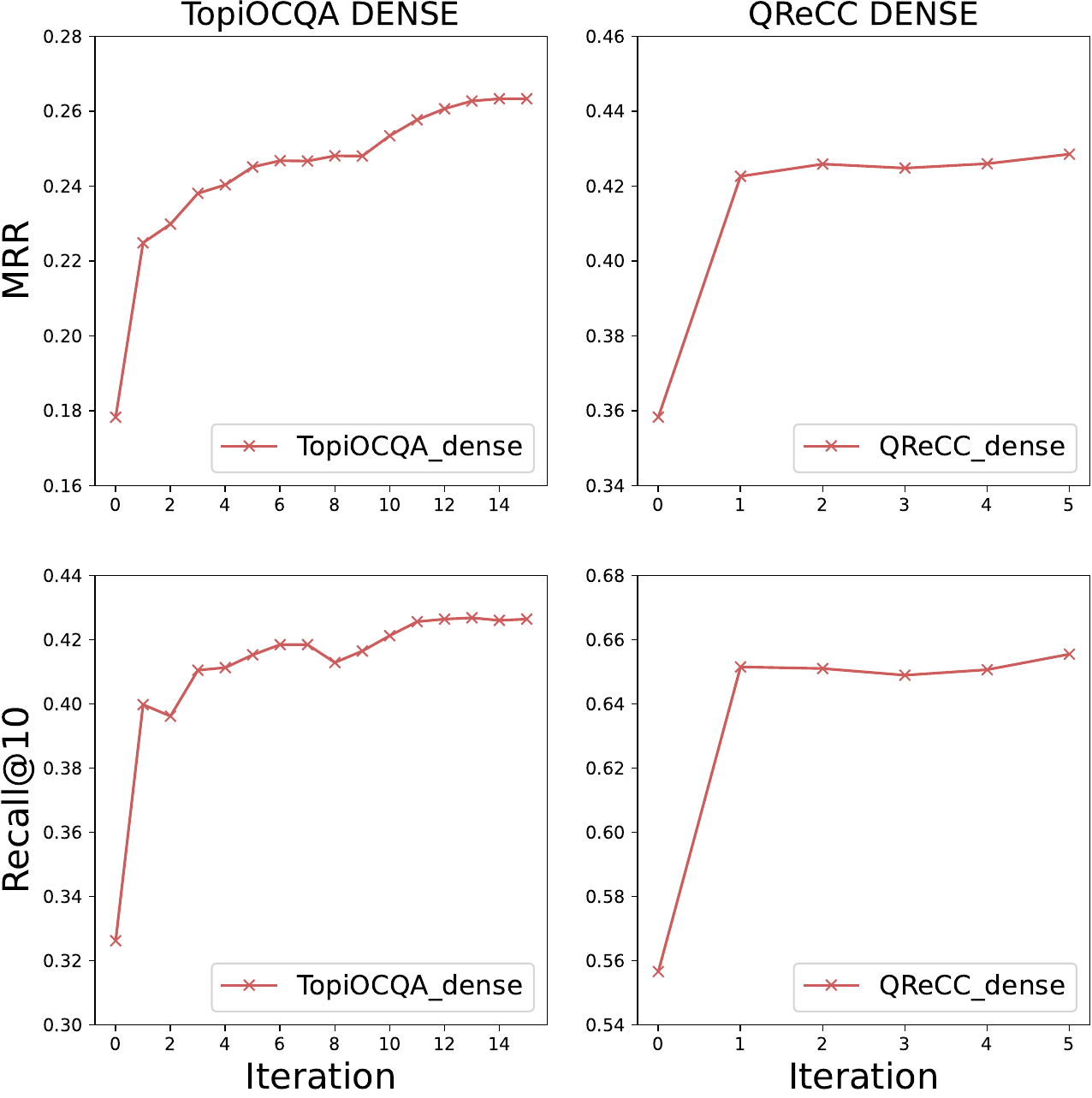}
\caption{IterCQR dense retrieval performance on TopiOCQA and QReCC datasets for each iteration.}
\label{fig:query_performance}
\vspace{-3mm}
\end{figure}

We show the retrieval performance of the TopiOCQA-trained model for each iteration in Figure \ref{fig:query_performance}.
In both the TopiOCQA and QReCC datasets, there is a notable improvement in the MRR and Recall@10 metric as the iterations progress.
In particular, applying the MBR training algorithm in iteration 1 significantly enhances performance.
This outcome suggests that the IterCQR effectively explored the query space, leading to significant improvements in retrieval performance.

\subsection{Effect on Query}
\begin{figure}
\centering
\includegraphics[width=\columnwidth]{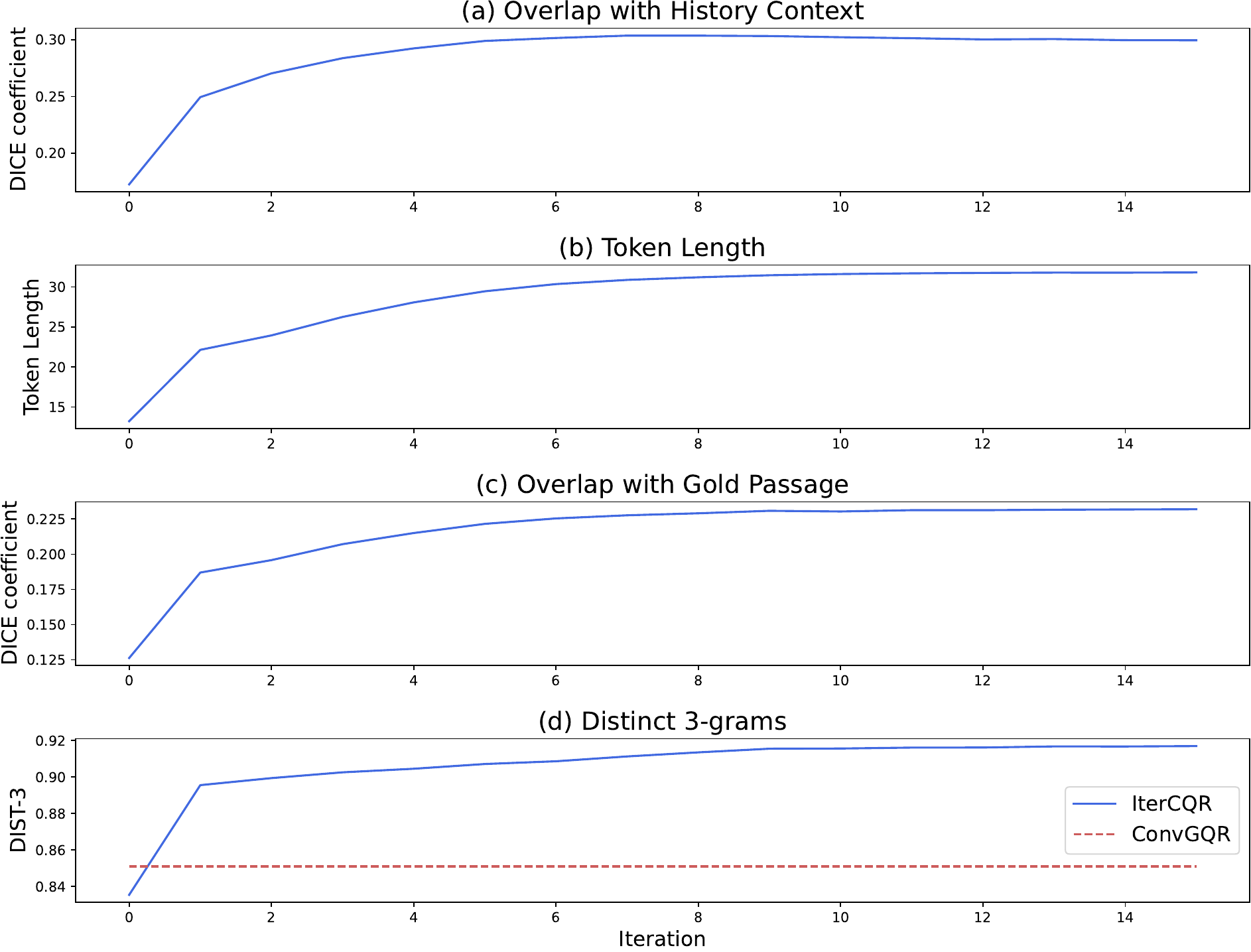}
\caption{Effect of iterative setting on queries. Overlapping tokens in (a) and (c) is shown by the  Sørensen-Dice coefficient, (b) is reported by the average token length of the reformulated queries, and (d) represents the proportion of distinct 3-gram tokens per query. All results are derived from the TopiOCQA test set.}
\vspace{-5mm}
\label{fig:query_overlap}
\end{figure}

We analyze the characteristics of the reformulated queries by IterCQR for each step trained with TopiOCQA and present the results in Figure \ref{fig:query_overlap}.
We utilize the Sørensen-Dice coefficient \cite{sorensen,dice} in Equation \ref{eq:dice} to measure the similarity of two strings.
\begin{equation}
    D(A,B)=\frac{2*|A\cap B|}{\left| A \right|+\left| B \right|}
\label{eq:dice}
\end{equation}
As the iterations progress, we observe a consistent increase in token overlap with historical context as shown in \ref{fig:query_overlap} (a).
Ultimately, this trend signifies that the IterCQR model progressively learns to extract information from historical context and integrates it into the current conversational query.

Furthermore, when examining the average query token length for each iteration, as shown in Figure \ref{fig:query_overlap} (b), evidently, the token length increases with each iteration.
These findings from Figure \ref{fig:query_overlap} (a) and (b) suggest that the IterCQR model learns to summarize the previous context.

To validate whether this summarization is helpful for retrieval, we measure the overlapping tokens with the gold passage using the Dice coefficient.
As depicted in Figure \ref{fig:query_overlap} (c), we consistently observe an increase in the token overlap with the gold passage. 
We also report the proportion of distinct 3-grams per query for each iteration in Figure \ref{fig:query_overlap} (d), demonstrating that the overlap of tokens with the gold passage is not merely due to simple keyword repetition.
We can even observe that IterCQR generates more diverse queries than those from ConvGQR, which utilizes potential answer expansion.

This implies that IterCQR's queries ultimately provide a stronger retrieval signal towards the gold passage, thereby contributing to better retrieval performance, as shown in Figure \ref{fig:query_performance}.
We provide generated queries by each iteration in Appendix \ref{iterative_query}.

Additionally, in Figure \ref{fig:query_overlap}, we consistently observe a sharp increase in iteration that employ MBR training method, specifically iteration 1.
This sharp increase aligns with the pattern of sharp increases in the retrieval performance in Figure \ref{fig:query_performance}. 
This pattern shows the effectiveness of the MBR training algorithm for exploration.

\section{Comparison between MBR and SCST}
\begin{table}[th]
\centering
\resizebox{\columnwidth}{!}{%
\begin{tabular}{cccccc}
\toprule[1.5pt]

\multirow{2}{*}{ \textbf{Type}} & \multirow{2}{*}{\textbf{Method}} & \multicolumn{4}{c}{TopiOCQA} \\
\cline{3-6}
& & MRR & NDCG@3 &R@10 & R@100\\
\hline
\multirow{3}{*}{Dense} 

& SCST-20 &20.6  & 19.2  & 35.6 & 55.4  \\
& SCST-50 &19.9  & 19.0 & 34.6 & 52.6  \\
& MBR & \textbf{22.5} & \textbf{21.1} & \textbf{40.0} & \textbf{58.9}    \\
\hline

\multirow{3}{*}{Sparse} 
& SCST-20 & 13.9&  11.2 &  27.2&  51.5   \\
& SCST-50 & 14.4&  12.6 &  26.9&  50.5   \\
& MBR & \textbf{16.4} & \textbf{14.8} & \textbf{30.2} & \textbf{54.8}   \\
\bottomrule[1.5pt]
\end{tabular}%
}
  \caption{ Comparison between SCST and MBR training algorithm. In the case of SCST, sampling was employed during candidate generation while MBR utilized beam search. We report the results considering two variations in top-k sampling: top-k=20 and 50. }
  \label{table:scst}
\end{table}

In this section, we compare the Self-Critical Sequence Training (SCST) utilized in \textsc{Con}QRR with our MBR training in IterCQR.
In the case of the SCST, sampling was employed during candidate generation, and we report two variations in top-k sampling: top-k=20 and top-k=50.
The outcomes presented in Table \ref{table:scst} indicate the performance of \(M_1\), trained using the MBR training algorithm and the SCST algorithm, starting from the same initial model \(M_0\). 
Evidently, MBR is a more effective algorithm for CQR, outperforming both dense and sparse retrievers for all evaluated metrics.
 \label{analysis}
\section{Conclusion}
In this work, we propose IterCQR, a methodology that iteratively improves CQR model without relying on human rewrites. 
IterCQR leverages retrieval signals when training CQR model, which provides retriever-friendly guidance for CQR.
We demonstrate the effectiveness of IterCQR through state-of-the-art performance on the QReCC and TopiOCQA datasets. 
In addition, the experimental results indicate that IterCQR learns to summarize previous contextual history, which leads to improved retrieval performance as the iteration progresses.
Furthermore, IterCQR exhibits superior performance in challenging settings such as generalization on unseen datasets and low-resource setting.
\section*{Limitations}
IterCQR employs the LLM, specifically \texttt{gpt-3.5-turbo}, to create the rewritten queries in initial dataset $D_0$ for training CQR model.
However, utilizing LLM to generate a rewrite requires inference costs. 
Furthermore, the IterCQR initial performance relies on the LLM's performance.
Still, we show that IterCQR can maintain its retrieval performance using only 50\% of the entire dataset, which could improve data efficiency and save LLM inference costs.

Since IterCQR generates $n$ candidates for each training instance, it necessitates larger storage capacity.
Additionally, the iterative framework can lead to relatively longer training times, though it does not require additional cost in the inference time.

IterCQR leverages dense embedding information as a reward term. 
Consequently, as the iterative learning process continues, it becomes increasingly optimized for dense retriever performance. 
However, the dense reward signal may not consistently enhance retrieval performance for sparse retrievers.
Nonetheless, it is worth highlighting that IterCQR outperforms powerful baseline methods in terms of sparse retrieval performance, despite using dense retrieval reward.

\section*{Ethical Statement}
We conducted experiments utilizing publicly available datasets, all of which are in English.
Our CQR model tends to generate summary expansions of the previous dialogue history. These expansion terms are dependent on the contextual history of the dialogue.
Therefore, if there is bias or inappropriate statements in the previous history context, the generated queries may also potentially contain such information.
\section*{Acknowledgements}
This work was supported by LG AI Research. 
This work was partly supported by Institute of Information \& communications Technology Planning \& Evaluation (IITP) grant funded by the Korea government(MSIT) [NO.2021-0-01343, Artificial Intelligence Graduate School Program (Seoul National University) \& NO.2022-0-00184, Development and Study of AI Technologies to Inexpensively Conform to Evolving Policy on Ethics \& NO.2021-0-02068, Artificial Intelligence Innovation Hub (Artificial Intelligence Institute, Seoul National University)], the BK21 FOUR program of the Education and Research Program for Future ICT Pioneers, Seoul National University in 2024.
K. Jung is with ASRI, Seoul National University, Korea.
The Institute of Engineering Research at Seoul National University provided research facilities for this work.

\bibliography{acl}
\clearpage
\appendix
\section{Implementation Details} \label{implementation}
In this work, we evaluate IterCQR on widely-used conversational search datasets.
TopiOCQA dataset consists of 3,920 conversations with average of 13 question-answer turns for each conversation.
For the TopiOCQA dataset, IterCQR is trained for 5 epochs for the initial \(M_0\) model, 2 epochs for MBR, and 5 epochs for the rest of the Top-1 candidate selection. The TopiOCQA-trained model is trained over 15 iterations.

QReCC dataset includes 13.6k conversations and each conversation consists of 6 turns in average.
For the QReCC dataset, IterCQR is trained for 10 epochs for the \(M_0\) model, MBR training for 2 epochs, and 5 epochs for Top-1 candidate selection. 
The final results of the QReCC-trained model are obtained after 5 iterations. 
For the experiments, we report the result with a single run of IterCQR, because it is costly to generate multiple initial dataset with LLM. 
All experiments are conducted using a single Nvidia A6000 GPU.
Training time differs by the training method: MBR training requires about 5 hours for one epoch, and Top-1 candidate selection requires about 40 minutes. 
We utilize T5-base as a backbone of the CQR model, which consists of 220M trainable parameters.


\section{Topic-shift Analysis}
CQA differs from standard Question Answering (QA) because of its interactive and conversational nature, introducing the concept of topic-shift during the conversation.
This topic-shift phenomenon has been recognized as a challenging aspect for models in various studies \cite{topiocqa, conqrr}.
To assess the capability of IterCQR to handle turns with topic shifts, we conducted an analysis of two distinct criteria associated with topic shifts.
\begin{figure}[t]
\includegraphics[width=\columnwidth]{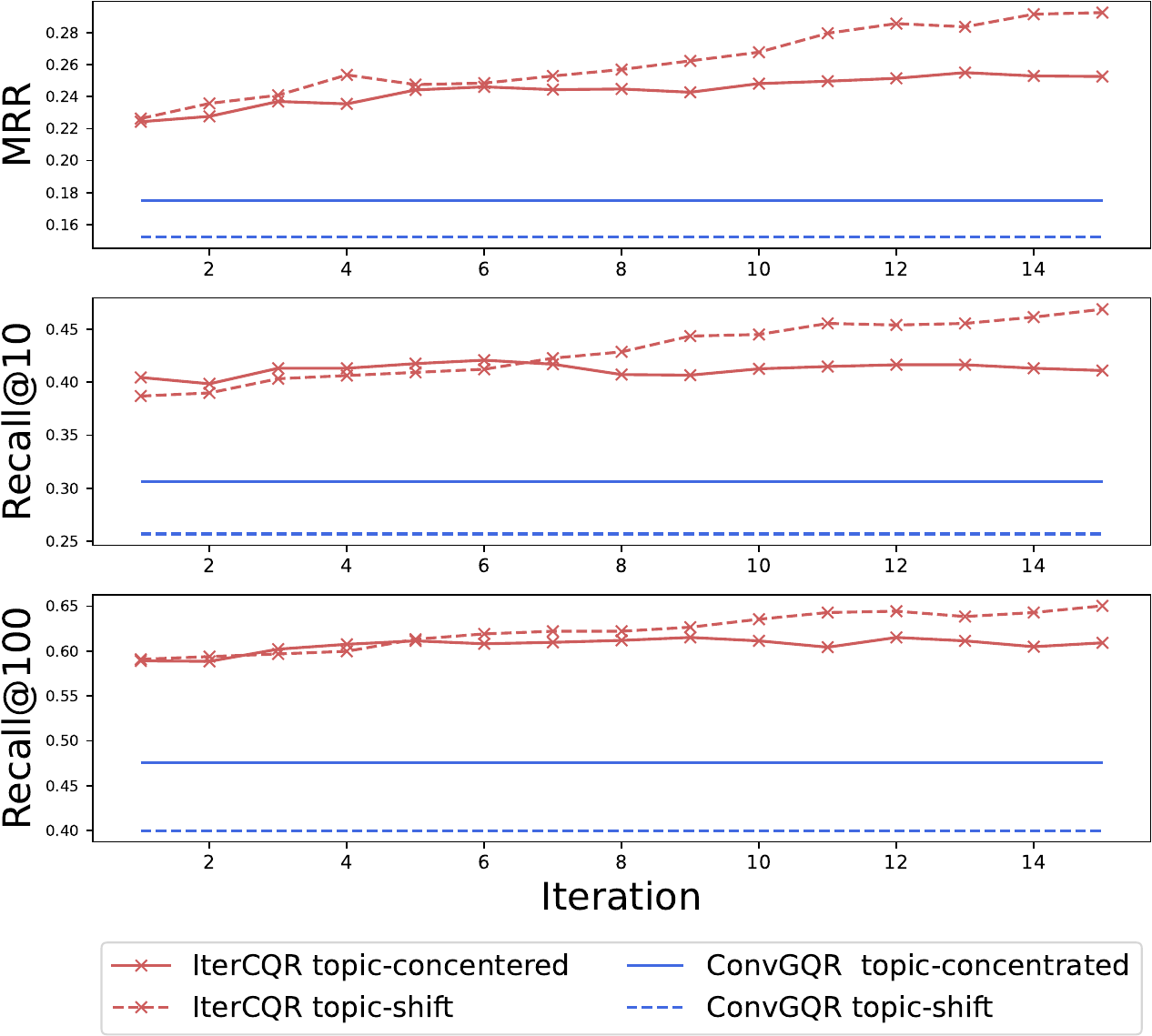}
\caption{ Performance of TopiOCQA-trained IterCQR on both topic-shifted and topic-concentrated examples. The TopiOCQA test set was divided by topic-shifted and topic-concentrated samples based on the topic label from the dataset. We report the retrieval performance of each iteration with metrics MRR, Recall@10, and Recall@100, respectively.}
\label{graph:topic_shift}
\end{figure}

%
\begin{table*}[t]
\centering
\normalsize
\resizebox{\textwidth}{!}{%
\begin{tabular}{@{\extracolsep{5pt}}ll|cccccc@{}}
\toprule[1.5pt]
             &      & \multicolumn{3}{c}{\textbf{Topic-Concentrated}} & \multicolumn{3}{c}{\textbf{Topic-Shifted}}     \\
\cline{3-5} \cline{6-8}
\textbf{Model} & \textbf{IR} & \textbf{MRR} & \textbf{Recall@10} & \textbf{Recall@100} & \textbf{MRR} & \textbf{Recall@10} & \textbf{Recall@100} \\ \hline
T5QR          & BM25 & 0.352           & 54.4          & 84.0          & \textbf{0.252}         & 45.1          & 79.1          \\
\textsc{Con}QRR(mix)   & BM25 & 0.419           & 63.1          & \textbf{91.2} & \textbf{0.252}         & 45.9          & \textbf{82.1} \\
\textsc{Con}QRR(RL)    & BM25 & 0.444           & 66.2          & 90.3          & 0.233          & 44.5          & 78.4          \\
IterCQR        & BM25 & \textbf{0.544}  & \textbf{72.4} & 89.7          & 0.249 & \textbf{49.7} & 77.7         \\ \hline
Human Rewrite & BM25 & 0.440           & 66.7          & 98.8          & 0.318          & 56.7          & 98.4          \\ 
\bottomrule[1.5pt]
\end{tabular}%
}
\caption{ Performance of QReCC-trained IterCQR on topic-concentrated and topic-shifted samples. In this experiment, topic shift was determined by gold passage ID in the QReCC dataset. For a fair comparison with \textsc{Con}QRR, we only report the performances on sparse retrieval.}
\label{table:topic-shift-bypid}
\end{table*}
\subsection{ Topic-shift by Topic Label}
We divide the TopiOCQA dataset into two categories: topic-shift and topic-concentrated, based on the assigned topic labels from the dataset.
In this categorization, we determine a topic-shifted instance if the topic label in the current turn differs from the topic label from the immediately preceding turn.
According to this criterion, the dataset is divided as topic-concentrated for 73\% , and topic-shift for 27\%.

We evaluate the performance of the TopiOCQA-trained IterCQR with a dense retriever in addressing topic-shift scenarios as shown in Figure \ref{graph:topic_shift}.
We observe a consistent improvement in performance across iterations for both topic-shifted and topic-concentrated scenarios.
Comparing our results with the performance of ConvGQR that we have reproduced, clearly, IterCQR outperforms ConvGQR in both topic-concentrated and topic-shifted scenarios, presenting our model's superior performance. 
ConvGQR performs better with topic-concentrated samples than with topic-shift instances, proving that topic-shifted turns are more difficult for the retriever.
However, with the MRR metric, IterCQR performs better on topic-shifted instances than on topic-concentrated cases.
In addition, in terms of Recall@10 and Recall@100, IterCQR initially shows a better performance on topic-concentrated cases; however, as the iterations progress, topic-shifted cases surpass that in the topic-concentrated cases.
This observation highlights the significant influence of IterCQR's cosine similarity reward based on dense representation, emphasizing performance improvement through iterative reformulation.

\subsection{Topic-shift by Gold Passage ID}
For the second criterion of topic-shift, we divide the QReCC test set based on the gold passage IDs within the dataset.
In this setting, if the gold passage ID of the current turn does not appear in any preceding turn within the same conversational session, a topic-shift is considered to have occurred.
According to this criterion, topic-concentrated instances account for 30\% of the dataset, whereas topic-shifted samples constitute 70\%.

Table \ref{table:topic-shift-bypid} provides a comparison of the results of topic-shifted cases determined by gold passage IDs. 
The scores reported in the \textsc{Con}QRR paper are presented in Table \ref{table:topic-shift-bypid}.
Note that it is hard to fairly compare IterCQR and \textsc{Con}QRR on dense retriever performance because \textsc{Con}QRR used DualEncoder instead of ANCE.
Hence, in this experiment, we compare the results on the sparse retriever.

The results indicate that IterCQR outperforms \textsc{Con}QRR across all instances, in both the topic-concentrated and topic-shifted scenarios, particularly in terms of the MRR metric. 
Especially, MRR scores of IterCQR are notably superior, even surpassing the human rewrite performance. 
IterCQR demonstrates comparable performance to \textsc{Con}QRR in addressing topic-shifted instances.
However, overall, human rewrites outperform all the models, demonstrating exceptional robustness in topic-shift scenarios.

\section{IterCQR with Other Expansions}

\begin{table}[t]
\centering
\resizebox{\columnwidth}{!}{%
\begin{tabular}{clccc}
\toprule[1.5pt]
\textbf{Dataset} & \textbf{Method} & \textbf{MRR} & \textbf{NDCG@3} & \textbf{R@10} \\ 
\hline
\multirow{2}{*}{\textbf{TopiOCQA}} & \textbf{IterCQR} & 0.263 & 0.251 & 42.6 \\
 & \textbf{IterCQR\small{+ expansion}} & 0.277 & 0.264 & 44.6 \\ 
\hline
\multirow{2}{*}{\textbf{QReCC}} & \textbf{IterCQR} & 0.429 & 0.402 & 65.5 \\
 & \textbf{IterCQR\small{+ expansion}} & 0.444 & 0.417 & 67.3 \\
\bottomrule[1.5pt]
\end{tabular}%
}
  \caption{ Performance with potential answer expansion. }
  \label{table:expansion}
\end{table}

In this experiment, we show that the factor of performance improvement in IterCQR is orthogonal to that of potential answer expansion introduced in ConvGQR\cite{convgqr}.
For TopiOCQA and QReCC, we concat the reformulated query generated by the IterCQR with an expansion term trained on gold answers of the dataset.
As shown in Table \ref{table:expansion}, with the potential answer expansion, there is an additional improvement in the retrieval performance, even though IterCQR has already achieved state-of-the-art performance on both datasets.

\section{Qualitative Analysis on Ablation Study} \label{ablation_query}
In section \ref{ablation}, we demonstrate that IterCQR exhibits superior performance when compared to solely using MBR training method and Top-1 candidate selection.
We provide comparative anlaysis on the queries generated by OnlyMBR and OnlyTop1 model with those generated by IterCQR in Table \ref{table:example_ablation}.
Reformulated queries produced by IterCQR contain essential information that effectively directs to the gold passage, consequently yielding significant improvements in retrieval performance.

\begin{table*}
\centering\resizebox{\textwidth}{!}{
\renewcommand{\arraystretch}{1.3}
\begin{tabular}{>{\hspace{0pt}}m{0.2\linewidth}>{\hspace{0pt}}m{0.9\linewidth}} 
\toprule [1.5pt]
\textbf{Type} & \multicolumn{1}{>{\arraybackslash\hspace{0pt}}m{0.873\linewidth}}{\textbf{Content}} \\ 
\hline
\textbf{Original Query} & what was \textcolor{red}{ the series }about? \\ 
\hdashline
\textbf{OnlyTop1 Query} & What is \textcolor{red}{Grey's Anatomy }about in the series "Grey's Anatomy" broadcasted by the American Broadcasting \\ 
\hdashline
\textbf{OnlyMBR Query~~} & What is "\textcolor{red}{Grey's Anatomy}", an American television series produced in 2005 and featuring several films including Melissa and George and Alex Proy \\ 
\hdashline
\textbf{IterCQR Query} & What is \textcolor{red}{ Grey's Anatomy}, a television series premiered on \textbf{\textcolor{blue}{March 27, 2005}} by \textcolor{blue}{\textbf{the American Broadcasting Company (ABC)}}. \\ 
\hdashline
\textbf{Gold Passage} & Grey's Anatomy Introduction \textcolor{red}{ Grey's Anatomy} is an American medical drama television series that premiered on \textcolor{blue}{\textbf{ March 27, 2005}}, on \textbf{\textcolor{blue}{ the American Broadcasting Company (ABC) }}as a mid-season replacement. The fictional series focuses on the lives of surgical interns, residents, and attendings as they develop into seasoned doctors ... \\ 
\hline
\textbf{Original Query} & who directed \textcolor{red}{ it}? \\ 
\hdashline
\textbf{OnlyTop1 Query} & Who directed \textcolor{red}{Ride Me to Hell}, the episode "Ride Me to Hell," the third episode of the American animated television series "Ugly Americans," \\ 
\hdashline
\textbf{OnlyMBR Query~ ~~} & Who directed "\textcolor{red}{Ride Me to Hell}", an episode of American animated television series featuring teenage lead characters including John and Julie, and Ryan in 1992 and ending \\ 
\hdashline
\textbf{IterCQR Query} & Who directed \textcolor{red}{Ride Me to Hell,} the episode of the American animated television series "Ugly Americans", which aired on \textbf{\textcolor{blue}{July 14, 2011}} titled \\ 
\hdashline
\textbf{Gold Passage} & Ride Me to Hell Introduction "\textcolor{red}{Ride Me to Hell}" is the third episode of the of the American animated television series "Ugly Americans", and the seventeenth overall episode of the series. It originally aired on Comedy Central in the United States on \textbf{\textcolor{blue}{ July 14, 2011}}. In the episode, ... \\
\bottomrule [1.5pt]
\end{tabular}}
\caption{ Generated queries by TopiOCQA-trained onlyTop1, onlyMBR, and IterCQR. \textcolor{red}{Red} words represnt rewritten entity and \textcolor{blue}{blue} words show the summary expansion included only in IterCQR. }
\label{table:example_ablation}
\end{table*}

\section{Effect of Iterative Setting on Queries} \label{iterative_query}
In section \ref{analysis}, we empirically show that IterCQR learns to generate summary expansion of the preceding history context as the iteration progresses. 
In Table \ref{table:example_iteration}, we provide examples of generated queries at different intermediate iterations of IterCQR, specifically at iterations 0, 1, 5, and 15. 
Evidently, IterCQR progressively acquires the capacity to distill essential information from the previous dialogue context, resulting in a higher token overlap with the gold passage and retrieval performance.

\begin{table*}
\centering\resizebox{\textwidth}{!}{
\renewcommand{\arraystretch}{1.3}
\begin{tabular}{>{\hspace{0pt}}m{0.16\linewidth}>{\hspace{0pt}}m{0.894\linewidth}} 
\toprule [1.5pt]
\textbf{Type} &\textbf{Content} \\ 
\hline
\textbf{Original Query} & name the structural parts of \textcolor{red}{this}. \\
\hdashline
\textbf{IterCQR (0)} & What are the structural parts of the \textcolor{red}{Milky Way galaxy}? \\
\hdashline
\textbf{IterCQR (1)} & What are the structural components of the \textcolor{red}{Milky Way galaxy}, which contains the Solar System? \\
 \hdashline
\textbf{IterCQR (5)} & What are the structural components of the \textcolor{red}{Milky Way galaxy}, which contains ten billion white dwarfs, a billion neutron stars, and a \\
\hdashline
\textbf{IterCQR (15)} & What are the structural components of the \textcolor{red}{Milky Way}, which is the \textcolor{blue}{\textbf{second-largest galaxy in the Local Group with its stellar disk approximately in diameter}} \\
\hdashline
\textbf{ Gold Passage} & 'Milky Way Size and mass The Milky Way is the \textcolor{blue}{\textbf{second-largest galaxy in the Local Group}} (after the Andromeda Galaxy), \textcolor{blue}{\textbf{with its stellar disk approximately in diameter}} and, on average, approximately thick. The Milky Way is approximately 890 billion times the mass of the Sun. To compare the relative physical scale of the Milky Way, if the Solar System out to Neptune were the size of a US quarter (), the Milky Way would be approximately the size of the contiguous United States. There is a ring-like filament of stars rippling above and below the relatively flat galactic plane, wrapping around the Milky Way at a diameter of , which may be part of the Milky Way itself. \\ 
\hline
\textbf{Original Query} & where is \textcolor{red}{it} located? \\
\hdashline
\textbf{IterCQR (0)} & Where is the aforesaid administration located? \\
\hdashline
\textbf{IterCQR (1)} & Where was the \textcolor{red}{National Oceanic and Atmospheric Administration (NOAA)} located in 1970? \\
 \hdashline
\textbf{IterCQR (5)} & Where was the U.S. \textcolor{red}{National Oceanic and Atmospheric Administration (NOA)} located during its formation on October 3, 1970? \\
 \hdashline
\textbf{IterCQR (15)} & Where is the U.S. \textcolor{red}{National Oceanic and Atmospheric Administration (NOAA}),\textcolor{blue}{\textbf{ an American scientific agency}}, formed on October 3, 1970 \\
\hdashline
\textbf{Gold Passage} & National Oceanic and Atmospheric Administration History NOAA traces its history back to multiple \textcolor{blue}{\textbf{agencies}}, some of which were among the oldest in the federal government: The most direct predecessor of NOAA was the Environmental Science Services Administration (ESSA), into which several existing \textcolor{blue}{\textbf{scientific agencies}} such as the United States Coast and Geodetic Survey, the Weather Bureau and the uniformed Corps were absorbed in 1965. NOAA was established within the Department of Commerce via the Reorganization Plan No. 4 and formed on October 3, 1970, after U.S. President Richard Nixon proposed creating a new agency to serve a national need for "better protection of life and property from natural hazards… for a better understanding of the total environment… [and] for exploration and development leading to the intelligent use of our marine resources \\

\bottomrule [1.5pt]
\end{tabular}}
    \caption{ Queries generated by TopiOCQA-trained IterCQR on intermediate iterations. IterCQR($t$) represents the model on iteration $t$. \textcolor{red}{Red} words note rewritten entity and \textcolor{blue}{blue} words show the summary expansion included only in IterCQR. }
    \label{table:example_iteration}
\end{table*}
Furthermore, IterCQR is capable of fixing the reformulation errors that exist in the earlier iteration of the IterCQR model.
In Table \ref{table:main_results}, the retrieval performance of $M_0$ trained on LLM-generated initial dataset $D_0$ is far inferior compared to our final IterCQR model.
We illustrate instances in Table \ref{table:example_initial} which IterCQR mitigates the errors in $M_0$ as iteration progresses.

\begin{table*}
\centering\resizebox{\textwidth}{!}{
\renewcommand{\arraystretch}{1.3}
\begin{tabular}{>{\hspace{0pt}}m{0.16\linewidth}>{\hspace{0pt}}m{0.894\linewidth}} 
\toprule [1.5pt]
\textbf{Type} & \multicolumn{1}{>{\arraybackslash\hspace{0pt}}m{0.894\linewidth}}{\textbf{Content}} \\ 
\hline
\textbf{Original Query} & what was \textcolor{red}{ the series }about? \\ 
\hdashline
\textbf{IterCQR (0)} & What is the plot of\uline{ \textcolor[rgb]{1,0.647,0}{\textbf{the series "Dark City"}}}? \\ 
\hdashline
\textbf{IterCQR (1)} & What was \uline{\textcolor{blue}{\textbf{the series "Grey's Anatomy"}}} about primarily about sports programming primarily on weekend afternoons \\ 
\hdashline
\textbf{IterCQR (15)} & What is\uline{ \textcolor{blue}{\textbf{Grey's Anatomy}}}, a television series premiered on March 27, 2005 by the American Broadcasting Company (ABC). \\ 
\hdashline
\textbf{Gold Passage} & \textcolor{blue}{Grey's Anatomy} Introduction \textcolor{blue}{Grey's Anatomy} is an American medical drama television series that premiered on March 27, 2005, on the American Broadcasting Company (ABC) as a mid-season replacement. The fictional series focuses on the lives of surgical interns, residents, and attendings as they develop into seasoned doctors while balancing personal and professional relationships. The title is an allusion to "Gray's Anatomy", a classic human anatomy textbook first published in 1858 in London and written by Henry Gray. Shonda Rhimes developed the pilot and continues to write for the series. She is also one of the executive producers alongside Betsy Beers, Mark Gordon, Krista Vernoff, Rob Corn, Mark Wilding, and Allan Heinberg and recently Ellen Pompeo. \\
\bottomrule  [1.5pt]
\end{tabular}}
\caption{ IterCQR mitigates initial model's error. IterCQR($t$) represents the model on iteration $t$. \textcolor{red}{Red} words show the coreference from the original query, \textcolor[rgb]{1,0.647,0}{orange} words show the initial model $M_0$ error and \textcolor{blue}{blue} words show the correct rewritten entity by IterCQR.  }
\label{table:example_initial}
\end{table*}


\section{Queries Generated by IterCQR}
In this section, we present the queries generated by IterCQR on both the TopiOCQA and QReCC dataset in Table \ref{table:topiocqa_example} and Table \ref{table:qrecc_example2}, respectively. 
Note that the TopiOCQA dataset does not have a human rewrite; therefore, we demonstrate a reformulated query generated by a model trained on the QReCC dataset's human rewrite. 

\begin{table*}
\centering\resizebox{\textwidth}{!}{
\renewcommand{\arraystretch}{1.2}
\begin{tabular}{>{\hspace{0pt}}m{0.17\linewidth}>{\hspace{0pt}}m{0.8\linewidth}} 
\toprule [1.5pt]
\multicolumn{2}{c}{\textbf{TopiOCQA Dataset}} \\
\hline
\multirow{4}{*}{\textbf{Previous Turns}} & Query: What is the symbol of flag of ecuador? \\
 & Answer: It \textbf{ \textcolor[rgb]{0,0.392,0}{consists of horizontal bands of yellow (double width), blue and red}.} \\
 & Query:~ Who designed it? \\
 & Answer:~UNANSWERABLE \\
 \hline
\textbf{Original Query} & Does \textcolor{red}{ it} resemble any other flag? \\
\hdashline
\textbf{Human Rewrite} & Does \textcolor{red}{ flag of ecuador} resemble any other flag \\
\hdashline
\textbf{LLM Rewrite} & Does the \textcolor{red}{flag of Ecuador} resemble any other flag?\\
\hdashline
\textbf{IterCQR Query} & What are some other flags that resemble the \textcolor{red}{ flag of Ecuador}, which is \textbf{ \textcolor[rgb]{0,0.392,0}{consists of horizontal bands of yellow (double width), blue, and red~}} \\
\hdashline
\textbf{Gold Answer} & Yes, Colombia and Venezuela \\
\hdashline
\textbf{Gold Passage} &  \textcolor{red}{Flag of Ecuador} Introduction The national \textcolor{red}{ flag of Ecuador}, which \textbf{ \textcolor[rgb]{0,0.392,0}{consists of horizontal bands of yellow (double width), blue and red}}, was first adopted by law in 1835 and later on 26 September 1860. The design of the current flag was finalized in 1900 with the addition of the coat of arms in the center of the flag. Before using the \textbf{ \textcolor[rgb]{0,0.392,0}{yellow, blue and red}} tricolor, Ecuador used white and blue flags that contained stars for each province of the country. The design of the flag is very similar to those of \textbf{\uline{Colombia and Venezuela}}, which are also former constituent territories of Gran Colombia. \\
\bottomrule [1.5pt]
\end{tabular}}
\caption{Reformulated Queries by IterCQR on TopiOCQA dataset.The \textcolor[rgb]{0,0.384,0}{green} and \textcolor{red}{red} words stand for \textcolor[rgb]{0,0.384,0}{overlap with previous context} and \textcolor{red}{rewritten entity}. \uline{Underlined words} notate the content that contains the gold answer for the given query. Note that human rewrite in the TopiOCQA dataset refers to the output of the model trained on the QReCC dataset's human oracle. }
\label{table:topiocqa_example}
\end{table*}
\begin{table*}
\centering\small\resizebox{\textwidth}{!}{
\renewcommand{\arraystretch}{1.4}
\begin{tabular}{>{\hspace{0pt}}m{0.17\linewidth}>{\hspace{0pt}}m{0.8\linewidth}} 
\toprule [1.5pt]
\multicolumn{2}{c}{QReCC Dataset} \\
\hline
\multirow{4}{*}{\hspace{0pt}\textbf{Previous Turns}} & Query: What is the role of work cover nsw \\
 & Answer: The agency \textcolor[rgb]{0,0.384,0}{\textbf{WorkCover}} NSW creates \textcolor[rgb]{0,0.384,0}{\textbf{regulations to promote productive, healthy and safe workplaces for workers in New South Wales}}. \\
 & Query: What else does the agency do \\
 & Answer: The agency WorkCover NSW created \textcolor[rgb]{0,0.384,0}{\textbf{regulations}} for employers too. \\ 
\hline
\textbf{Original Query} & Was there any controversy with \textcolor{red}{ the agency} \\
\hdashline
\textbf{Human Rewrite} & Was there any controversy with \textcolor{red}{nsw} \\
\hdashline
\textbf{LLM Rewrite} & Has the agency WorkCover \textcolor{red}{NSW} faced any controversy? \\
\hdashline
\textbf{IterCQR Query} & Was there any controversy surrounding the agency \textcolor[rgb]{0,0.384,0}{\textbf{WorkCover} }\textcolor{red}{NSW}'s \textcolor[rgb]{0,0.384,0}{\textbf{regulations} }to promote productive, \textcolor[rgb]{0,0.384,0}{\textbf{healthy and safe workplaces}} for workers in \textcolor[rgb]{0,0.384,0}{\textbf{New South Wales.}} \\
\hdashline
\textbf{Gold Answer} & In December 2005, the Independent Commission Against Corruption found that 23 NSW WorkCover employees had issued false certificates of competency. \\
\hdashline
\textbf{Gold Passage} & Dangerous Goods (Gas Installations) \textbf{\textcolor[rgb]{0,0.392,0}{Regulation}} 1998 (\textcolor{red}{NSW}) Dangerous Goods (Road and Rail Transport) \textcolor[rgb]{0,0.392,0}{\textbf{Regulation }}2009 (NSW) Explosives \textbf{\textcolor[rgb]{0,0.392,0}{Regulation}} 2005 (\textcolor{red}{NSW}) Occupational~\textcolor[rgb]{0,0.384,0}{\textbf{\textbf{Health}}}~and\textbf{\textbf{~\textcolor[rgb]{0,0.384,0}{Safety}~\textcolor[rgb]{0,0.384,0}{Regulation~}}}2001 (\textcolor{red}{NSW}) Sporting Injuries Insurance \textcolor[rgb]{0,0.392,0}{\textbf{Regulation }}2009 (\textcolor{red}{NSW}) Workers Compensation (Bush Fire, Emergency and Rescue Services) \textcolor[rgb]{0,0.392,0}{\textbf{Regulation }}2007 (\textcolor{red}{NSW}) Workers' Compensation (Dust Diseases) \textbf{\textcolor[rgb]{0,0.392,0}{Regulation}} 2008 (\textcolor{red}{NSW}) Workers Compensation \textcolor[rgb]{0,0.392,0}{\textbf{Regulation}} 2010 (\textcolor{red}{NSW}) ...\textbf{\textbf{~\uline{In December 2005, the Independent Commission Against Corruption found that 23~\textcolor[rgb]{0,0.384,0}{WorkCover}~employees had issued false certificates of competency}}}, which ICAC states significantly undermined~\textbf{\textbf{\textcolor[rgb]{0,0.384,0}{workplace}~}}safety on building sites. ... [18] In 2002, a\textbf{\textbf{~\textcolor[rgb]{0,0.384,0}{New South Wales}}}~parliamentary committee criticized the~\textbf{\textbf{\textcolor[rgb]{0,0.384,0}{WorkCover}}}~Authority. [19] \\

\hline
\textbf{Original Query} & When was \textcolor{red}{the agency} formed \\ 
\hdashline
\textbf{Human Rewrite} & when was \textcolor{red}{nsw} formed \\ 
\hdashline
\textbf{LLM Rewrite} & When was WorkCover \textcolor{red}{NSW} formed? \\
\hdashline
\textbf{IterCQR Query} & When was the \textcolor{red}{NSW} WorkCover agency formed to \textcolor[rgb]{0,0.384,0}{\textbf{promote productive, healthy and safe workplaces for workers in New South Wales.}} \\ 
\hdashline
\textbf{Gold Answer} & The WorkCover Authority of New South Wales was a New South Wales Government agency established in 1989. \\ 
\hdashline
\textbf{Gold Passage} & \textbf{\textbf{\textcolor[rgb]{0,0.384,0}{WorkCover}}}~Authority of~\textbf{\textbf{\textcolor[rgb]{0,0.384,0}{New South Wales}}}~- Wikipedia CentralNotice~\textbf{\textbf{\textcolor[rgb]{0,0.384,0}{WorkCover}}}~Authority of~\textbf{\textbf{\textcolor[rgb]{0,0.384,0}{New South Wales}}}~From Wikipedia, the free encyclopedia Jump to navigation Jump to search Authority of~\textcolor[rgb]{0,0.384,0}{\textbf{\textbf{New South Wales}}}~Statutory authority overview \textbf{\uline{Formed 1989}} Dissolved 2015 Jurisdiction~\textbf{\textbf{\textcolor[rgb]{0,0.384,0}{New South Wales}}}~Parent Statutory authority Department of Finance and Services Key documents Safety, Return to Work and Support Board Act, 2012 (\textcolor{red}{NSW}) Work Health and Safety Act, 2011 (\textcolor{red}{NSW}) Website~\textcolor[rgb]{0,0.384,0}{\textbf{\textbf{workcover}}}~.\textcolor{red}{nsw} .gov .au The\textbf{\textbf{\textcolor[rgb]{0,0.384,0}{~WorkCover}}}~Authority of~\textcolor[rgb]{0,0.384,0}{\textbf{\textbf{New South Wales}}}~or~\textcolor[rgb]{0,0.384,0}{\textbf{\textbf{WorkCover}}}~\textcolor{red}{NSW} is a~\textcolor[rgb]{0,0.384,0}{\textbf{\textbf{New South Wales~}}}Government agency established in 1989. The agency creates regulations to~\textbf{\textbf{\textcolor[rgb]{0,0.384,0}{promote productive, healthy and safe~workplaces for workers}~}}and employers in~\textcolor[rgb]{0,0.384,0}{\textbf{\textbf{New South Wales}}.}~[1] The agency formed part of the Safety, Return to Work and Support Division established pursuant to the Safety, Return to Work and Support Board Act, 2012 (\textcolor{red}{NSW}). On 1 September 2015,~\textbf{\textbf{\textcolor[rgb]{0,0.384,0}{WorkCover~}}}\textcolor{red}{NSW} was replaced by three new entities – Insurance and Care \textcolor{red}{NSW }(icare),...The information below pertains to the former~\textcolor[rgb]{0,0.384,0}{\textbf{\textbf{WorkCover}}}~\textcolor{red}{NSW}.~\textcolor[rgb]{0,0.384,0}{\textbf{\textbf{WorkCover}}}~\textcolor{red}{NSW} no longer exists, however its functions have been split between the aforementioned newly created agencies. \\
\bottomrule [1.5pt]
\end{tabular}}
\caption{Reformulated Queries by IterCQR on QReCC dataset. The \textcolor[rgb]{0,0.384,0}{green} and \textcolor{red}{red} words stand for \textcolor[rgb]{0,0.384,0}{overlap with previous context} and \textcolor{red}{rewritten entity}. \uline{Underlined words} notate the content that contains the gold answer for the given query.  }
\label{table:qrecc_example2}
\end{table*}

\section{Prompt Used for Dataset initilaization} \label{prompt}
We report the prompt used for generating $D_0$ using \texttt{gpt-3.5-turbo}.
We use the instruction: \textit{"This is a part of conversational question answering. Rewrite the current query as a stand-alone question based on the previous conversation so that it could be context-independent."}. We concat three preceding original queries and answers for the dialogue history context and the current query as the input for the LLM. 

\section{Failure Case} 
We demonstrate the failure case of IterCQR. Training the IterCQR with cosine similarity with the gold passage leads the model to increase the keyword overlap with the gold passage. However, this could also result in the repetition of the keyword in the reformulated query, as shown in Table \ref{table:failure}. 
\begin{table*}
\centering\resizebox{\textwidth}{!}{
\renewcommand{\arraystretch}{1.2}
\begin{tabular}{>{\hspace{0pt}}m{0.17\linewidth}>{\hspace{0pt}}m{0.9\linewidth}} 
\toprule [1.5pt]
\textbf{Type} & {\textbf{Content}} \\ 
\hline
\multirow{2}{*}{\hspace{0pt}\textbf{Previous Turns}} & Query: What is the meaning of the song alejandro \\
 & Answer: The song bids farewell to her lovers. \\ 
\hdashline
\textbf{Original Query} & whose song is it? \\ 
\hdashline
\textbf{Gold Answer} & Lady Gaga \\ 
\hdashline
\textbf{IterCQR Query} & Who released the song "\textcolor{red}{Alejandro}" by \textcolor{red}{Alejandro}, a song that bids farewell to her lovers and bids far \\ 
\hdashline
\textbf{Gold Passage} & \textcolor{red}{Alejandro} (song) Introduction "\textcolor{red}{Alejandro}" is a song by American singer \uline{Lady Gaga}. It was released as the third single from her third EP, "The Fame Monster" (2009). Co-written and produced by Gaga and Nadir "RedOne" Khayat, it was inspired by her "Fear of Men Monster". The singer bids farewell to her lovers over mid-tempo synth-pop music with a Europop beat. Contemporary critics predominantly gave "\textcolor{red}{Alejandro}" positive reviews and noted that it takes influence from the pop acts ABBA and Ace of Base. The song charted in the United Kingdom and Hungary due to digital sales following the album\textbackslash{}'s release. Upon release, "\textcolor{red}{Alejandro}" charted again in the United Kingdom as well as in Australia, Canada, New Zealand, Sweden, and the United States while topping the Czech, Finnish, Mexican, Venezuelan, Polish, Russian, Bulgarian, and Romanian charts. \\
\bottomrule [1.5pt]
\end{tabular}}
\caption{ Failure case of TopiOCQA-trained IterCQR. In this case, the IterCQR query includes the keyword "Alejandro" repeatedly, deviating from the previous dialogue context.}
\label{table:failure}
\end{table*}
\end{document}